\documentclass[jpcfk,twocolumn,groupeaddress,prb]{revtex4}

\usepackage{graphicx}

\usepackage{amsfonts, amsmath, amssymb,latexsym}

\usepackage{dcolumn}

\usepackage{bm}

\begin{document}

\title{Concentration dependence of the Flory-Huggins interaction
parameter in aqueous solutions of capped PEO chains}

\author{M. I. Chaudhari, L. R. Pratt} 
\affiliation{Department of Chemical and Biomolecular Engineering, Tulane University, New Orleans, LA 70118}
\author{M. E. Paulaitis} 
\affiliation{Department of Chemical and Biomolecular Engineering, The Ohio State University,
Columbus, OH 43210}

\date{\today}

\begin{abstract}The dependence on volume fraction $\varphi$ of the Flory-Huggins
$\chi_{\mathrm{wp}}\left(\varphi\right)$ describing the free energy of mixing of
polymers in water is obtained by exploiting the connection of
$\chi_{\mathrm{wp}}\left(\varphi\right)$ to the chemical potential of the water,
for which quasi-chemical theory is satisfactory. We test this theoretical
approach with simulation data for aqueous solutions of capped PEO oligomers. For
CH$_3$(CH$_2$-O-CH$_2$)$_m$CH$_3$ ($m$=11),
$\chi_{\mathrm{wp}}\left(\varphi\right)$ depends strongly on $\varphi$,
consistent with experiment. These results identify coexisting water-rich and
water-poor solutions at $T$ = 300~K and $p$ = 1~atm. Direct observation of the
coexistence of these two solutions on simulation time scales supports that
prediction for the system studied. This approach directly provides the osmotic
pressures. The osmotic second virial coefficient for these chains is positive,
reflecting repulsive interactions between the chains in the water, a \emph{good}
solvent for these chains. \end{abstract}

\maketitle

\section{Introduction}
A classic element of polymer solution physics, 
the Flory-Huggins (FH) model,\cite{Flory:1953ty,Huggins:1941wb}
\begin{multline}
	\frac{\beta\Delta{G_{\mathrm{mix}}}}{n_\mathrm{w} + M n_\mathrm{p}} = 
	\varphi \ln\varphi + \frac{\left(1-\varphi\right)}{M}\ln\left(1-\varphi\right) 
	\\
	+ \varphi\left(1-\varphi\right)\chi_{\mathrm{wp}}~,
	\label{eq:1}
\end{multline}
describes the free energy of mixing of $n_\mathrm{p}$ moles of polymer liquid
with $n_\mathrm{w}$ moles of the water solvent; $\beta = 1/kT$, $\varphi$ is the
solvent volume fraction, $M = \bar{v}_\mathrm{p}/\bar{v}_\mathrm{w}$ (the ratio
of the molar volumes of the pure liquids) is the operational polymerization
index, and $\chi_{\mathrm{wp}}$ is the FH interaction coefficient. Here we study
the concentration dependence of $\chi_{\mathrm{wp}}$,  important for
mixing the dissimilar liquids  of water and chain molecules
that have a non-trivial aqueous solubility. The FH model is routinely adopted
for discussion of aqueous solutions of chain molecules of sub-polymeric
length.\cite{Sharp:1991tv,DeYoung:1990tv,Stillinger:1983kp} The study below
highlights direct access to the osmotic pressures of these solutions, and thus
can address long-standing research on biophysical hydration
forces.\cite{parsegian2011hydration}

Though the traditional statistical mechanical calculation\cite{lowHill} that
arrives at Eq.~\eqref{eq:1} is not compelling for aqueous materials, the FH
model captures two dominating points. Firstly, it identifies the volume fraction
$\varphi$ as the preferred concentration variable, associated with the physical
assumption that the excess volume of mixing vanishes. This step partially avoids
difficult statistical mechanical packing problems.\cite{Beck:2006tb} Secondly,
Eq.~\eqref{eq:1} captures the reduction of the chain molecule ideal entropy by
the factor $1/M$. The physical identification of the polymerization index $M$ as
$\bar{v}_\mathrm{p}/\bar{v}_\mathrm{w}$ as thermodynamically consistent as noted
below, but is a crude description of the molecular structure of polymers. With
these points recognized, however, the interaction contribution of
Eq.~\eqref{eq:1} can be regarded as an interpolation between the ends $\varphi =
0, 1$ of the composition range.

The simplest expectation \cite{lowHill,doi,Bae:1993uj,Beck:2006tb} for the
interaction parameter is
\begin{eqnarray}
 \chi_{\mathrm{wp}}       \propto  - \beta \left(
 a_{\mathrm{ww}}- \frac{ 2 a_{\mathrm{wp}}}{M}+ \frac{a_{\mathrm{pp}}}{M^2}   \right)
 \label{eq:naive}
\end{eqnarray}
where the parameters $a_{\eta \nu}$ gauge the strength of dispersion
interactions in van der Waals models of liquids.\cite{chandler1983van} That this
justification is implausible for aqueous solutions\cite{Shah:144508} underscores
the lack of a basic understanding of $ \chi_{\mathrm{wp}} $ for aqueous
solutions.

The simple temperature dependence of Eq.~\eqref{eq:naive} is a reasonable
starting point, but aqueous solutions exhibit alternative temperature
dependences of specific interest, hydrophobic effects.\cite{pohorille2012} More
troublesome, Eq.~\eqref{eq:naive} does not depend on concentration, though
experiments on the PEG/water system \cite{Bae:1993uj,Eliassi:1999wr} show
substantial concentration dependence. Beyond that difficulty, those results
exhibit a temperature trend opposite to Eq.~\eqref{eq:naive}, \emph{i.e.,}
stronger interactions at higher $T$ \ consistent with the classic folklore of
hydrophobic effects.\cite{Bae:1993uj} In contrast, when the solvent is methanol
\cite{ZafaraniMoattar:2006kr} the observed concentration dependence is less
strong, though non-trivial and trending with concentration in the opposite
direction from the aqueous solution results. The temperature dependences for the
methanol case is qualitatively consistent with the simple expectation of
Eq.~\eqref{eq:naive}. In further contrast, with ethanol as solvent
\cite{ZafaraniMoattar:2008tb} the observed concentration dependence is
distinctly modest.

\section{Theory}
These puzzles may be addressed by analyzing the chemical potential of
the  water,\cite{Bae:1993uj}
\begin{multline}
	\beta\Delta\mu^{\mathrm{(ex)}}_\mathrm{w} = 
	\left(1-\frac{1}{M}\right)\left(1-\varphi\right) \\
	+ 
	\frac{\partial\left(\varphi\chi_{\mathrm{wp}}\right)}{\partial\varphi}\left(1-\varphi\right)^2~,
\label{eq:2}
\end{multline}
where 
\begin{eqnarray}
\Delta\mu^{\mathrm{(ex)}}_\mathrm{w} = 
\mu^{\mathrm{(ex)}}_\mathrm{w}\left(\varphi,p, T\right) - 
\mu^{\mathrm{(ex)}}_\mathrm{w}\left(\varphi=1,p, T\right)~,
\label{eq:sstatedif}
\end{eqnarray}
the interaction (or \emph{excess}) contribution to the chemical potential of the
water, referenced to the pure liquid value.   The osmotic pressure $\pi$, \cite{lowHill}
\begin{eqnarray}
\beta \pi \bar{v}_\mathrm{w} = -\ln \varphi - 
\beta \Delta \mu^{\mathrm{(ex)}}_\mathrm{w}~.
\label{eq:pi}
\end{eqnarray}
provides further perspective on $\Delta\mu^{\mathrm{(ex)}}_\mathrm{w}$. Beyond
assuming that the excess volume of mixing vanishes, Eq.~\eqref{eq:pi} makes the
standard approximation that the solvent is incompressible.\cite{KO} To justify
the identification $M= \bar{v}_\mathrm{p}/\bar{v}_\mathrm{w}$ noted above, we
utilize Eq.~\eqref{eq:2}, and expand through $\left(1-\varphi\right)^2$ to
obtain
\begin{eqnarray}
\beta \pi V/n_\mathrm{p} \sim 1 + 
\left(\frac{n_\mathrm{p}}{V}\right)\bar{v}_\mathrm{w}M^2
\left( \frac{1}{2} - \hat{\chi}\right)~,
\label{eq:piB2}
\end{eqnarray}
with $\bar{v}_\mathrm{w}M^2 \left( \frac{1}{2} - \hat{\chi}\right) = B_2$ thus
the osmotic second virial coefficient. We adopt here the short-hand notation $
\hat\chi = \partial\left(\varphi\chi_{\mathrm{wp}}\right)/\partial\varphi.$ The
identification of $M$ in the contribution
$\left(1-\varphi\right)\left(1-\frac{1}{M}\right)$ thus leads to the proper
behavior in the ideal solution limit.

\begin{figure}
\begin{center}
\includegraphics[width=3.2in]{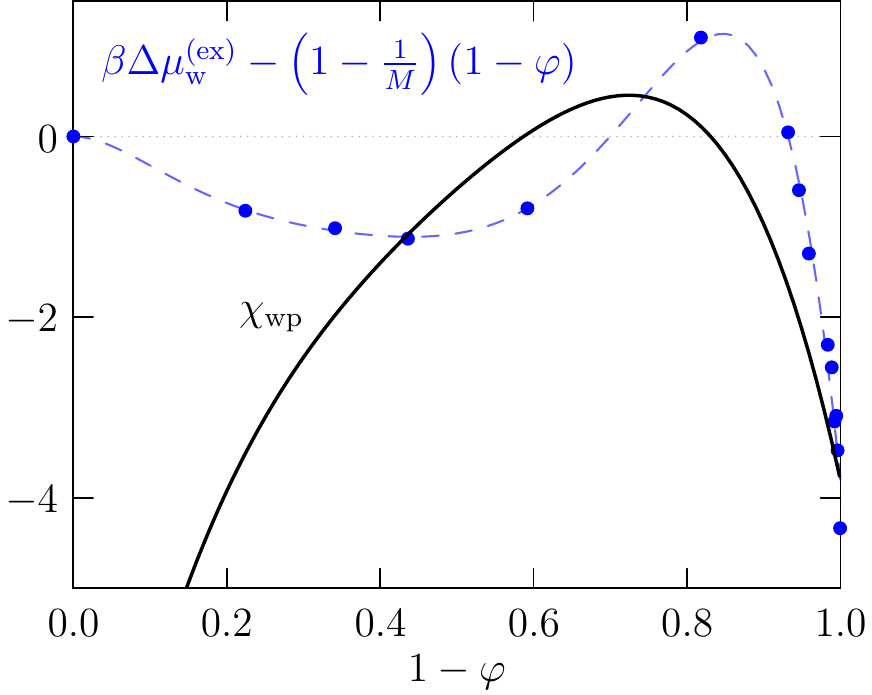}
\caption{Blue points and dashed curve: accumulated evaluation of
Eq.~\eqref{eq:2}. The solid curve is the implied $\chi_{\mathrm{WP}}
\left(\varphi\right)$.}
\label{fig:muxsV3}
\end{center}
\end{figure}

As suggested above, the intent of the FH model is that a
concentration-independent $\chi_{\mathrm{wp}}$ should describe the effects of
enthalpic interactions. Our less-committal analysis acquires practical
significance from recent development of molecular quasi-chemical theory for the
excess chemical potential of the water in aqueous
solutions.\cite{Shah:144508,Paliwal2006,Chempath:2009ws}
The central result  
\begin{multline}
	\beta\mu^{\mathrm{(ex)}}_\mathrm{w}\left(\varphi,p, T\right)  = -\ln p^{(0)}(n_{\lambda}=0) \\
	+ \ln\langle {e}^{\beta\varepsilon}\mid {n_{\lambda}=0}\rangle
	+ \ln p(n_\lambda = 0)
\label{eq:qct}
\end{multline} 	
is a physical description in terms of packing, outer-shell and chemical
contributions, a comprehensive extension of a van~der~Waals
picture.\cite{chandler1983van} The packing contribution is obtained from the
observed probability $p^{(0)}(n_\lambda=0)$ for successful random insertion of a
spherical cavity of radius $\lambda$ into the simulation cell. Similarly, the
chemical contribution is defined with the probability $p(n_\lambda=0)$ that a
water molecule in the system has zero neighbors within the
radius $\lambda$ of its O atom. The outer-shell contribution is a partition
function involving the binding energy $\varepsilon$, conditional on the
inner-shell being empty. The condition permits a
Gaussian statistical approximation,  
\begin{multline}
	\ln\langle {e}^{\beta\varepsilon}\mid {n_{\lambda}=0}\rangle
	\approx \beta\langle\varepsilon\mid{n_\lambda=0}\rangle \\
	+\beta^2\langle\delta\varepsilon^2\mid{n_\lambda=0\rangle}/2~,
\label{eq:5}
\end{multline}
involving the mean and variance of binding energies of molecules that have zero
neighbors within radius $\lambda$.

\begin{figure}
\begin{center}
\includegraphics[width=3.2in]{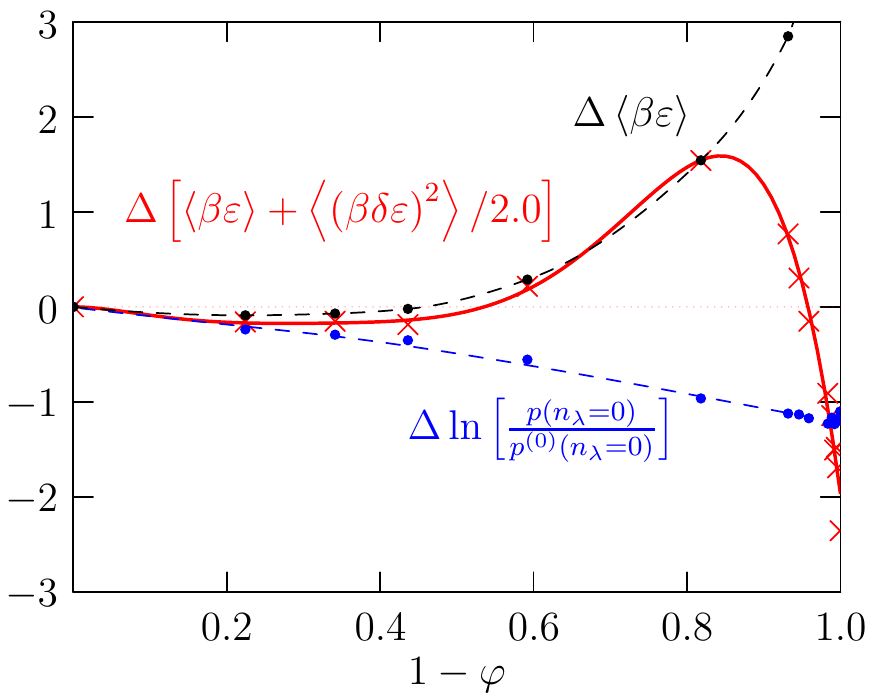}
\caption{`$\Delta$' indicates the difference from the pure solvent value,
\emph{i.e.,} $\Delta\left\langle \beta\varepsilon\right\rangle $ = $\left\langle
\beta\varepsilon\right\rangle \left(\varphi\right)$ - $\left\langle
\beta\varepsilon\right\rangle \left(\varphi=1\right)$. A water molecule loses
stabilizing outer-shell interactions through intermediate concentrations, and then
regains favorable outer-shell contributions on the solvent-poor side of the
concentration range. The dashed-blue curve shows  the combined packing and chemical
contributions.} 
\label{fig:muxspiecesV1}
\end{center}\end{figure}  

With this background, we evaluate
\begin{multline}
	\frac{\partial\left(\varphi\chi_{\mathrm{wp}}\right)}{\partial\varphi} \left(1-\varphi\right)^2=  
   \beta\Delta\mu^{\mathrm{(ex)}}_\mathrm{w} \\
   -\left(1-\frac{1}{M}\right)\left(1-\varphi\right)~.
\label{eq:2again}
\end{multline}
Representing then
\begin{eqnarray}
\frac{\partial\left(\varphi\chi_{\mathrm{wp}}\right)}{\partial\varphi}
= \sum_{n=0} c_n \left(1-\varphi\right)^n~,
\label{eq:sum}
\end{eqnarray}
and integrating, with $\varphi\chi_{\mathrm{wp}}=0$ at $\varphi =0$, we obtain
\begin{eqnarray}
\chi_{\mathrm{wp}} = \sum_{n=0} \frac{c_n}{\left(n+1\right)\varphi} \left\lbrack 1 - \left(1-\varphi\right)^{n+1} \right\rbrack~.
\label{eq:realchi}
\end{eqnarray}

\section{Results and Discussion} We thus analyze
$\chi_{\mathrm{WP}}\left(\varphi\right)$ for aqueous solutions for methyl-capped
PEO oligomers \cite{chaudhari_communication:_2010}
CH$_3$(CH$_2$-O-CH$_2$)$_{11}$CH$_3$ on the basis of accessible molecular
simulation data.\cite{chaudhari2014molecular} For this mixture we find $M$ = 27.6, with the excess volumes of
mixing similar to experimental results for similarly sized PEG
400:\cite{Eliassi:1999wr} negative and small, though slightly larger than the
comparable experimental case. The dielectric constant of these solution varies
linearly with solvent volume fraction $\varphi$.

Eq.~\eqref{eq:qct} is correct for any physical $\lambda$,
\cite{Paliwal2006} and we choose $\lambda = 0.29$~nm as a balance between
statistical and systematic accuracy. The Gaussian approximation will be more
accurate for larger $\lambda$. But the data set satisfying the condition
$n_\lambda=0$ gets smaller and the statistical accuracy is degraded with
increasing $\lambda$. The latter point becomes more serious at lower water
concentrations because fewer water molecules are present. Nevertheless, only the
difference Eq.~\eqref{eq:sstatedif} is required, so systematic errors should be
balanced to some extent. 

Composing Eq.~\eqref{eq:2} produces a structured dependence on $\varphi$
(FIG.~\ref{fig:muxsV3}). Extracting the individual quasi-chemical theory
contributions (FIG.~\ref{fig:muxspiecesV1}) shows that the distinctive variation
with $\varphi$ is due to the outer-shell (long-ranged) contributions: a water
molecule looses stabilizing outer-shell interaction partners through
intermediate concentrations, then regains favorable free energies through the
fluctuation contribution of the Gaussian formula (Eq.~\eqref{eq:5}). These
countervailing trends are not synchronous, so the net result is a non-monotonic
function of $\varphi$. 

\begin{figure}
\begin{center}
\includegraphics[width=3.2in]{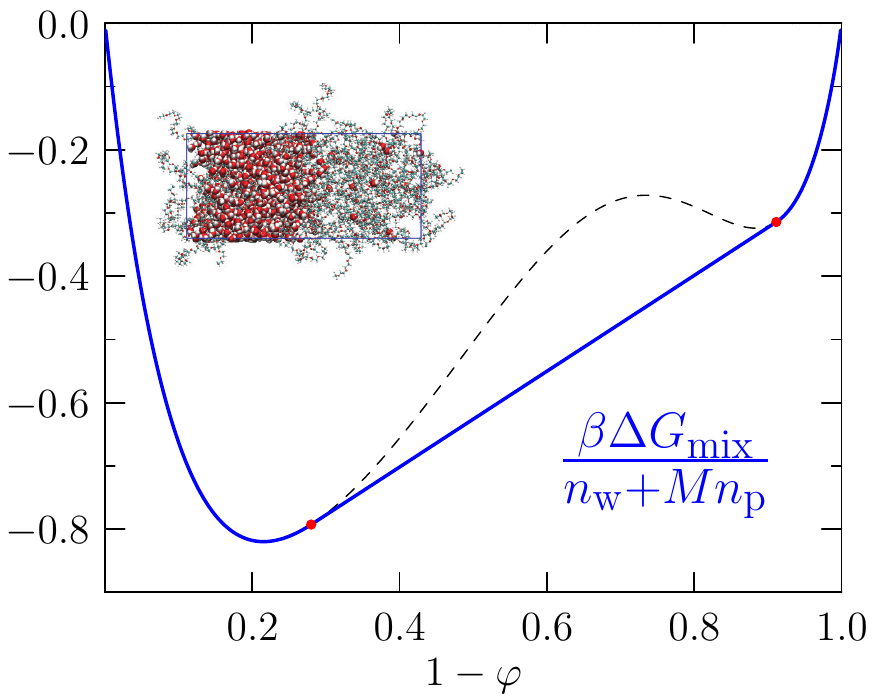}
\caption[Mixing free energy and double tangent construction]{Mixing free energy
and double tangent construction. INSET: Coexisting phases with chain-molecule
volume fraction $1 - \varphi \approx 0.34, 0.99$. This coexistence is
stable on the simulation time scale of 20~ns.} \label{fig:dGmixV3}
\end{center}
\end{figure} 

\begin{figure}
\begin{center}
\includegraphics[width=3.2in]{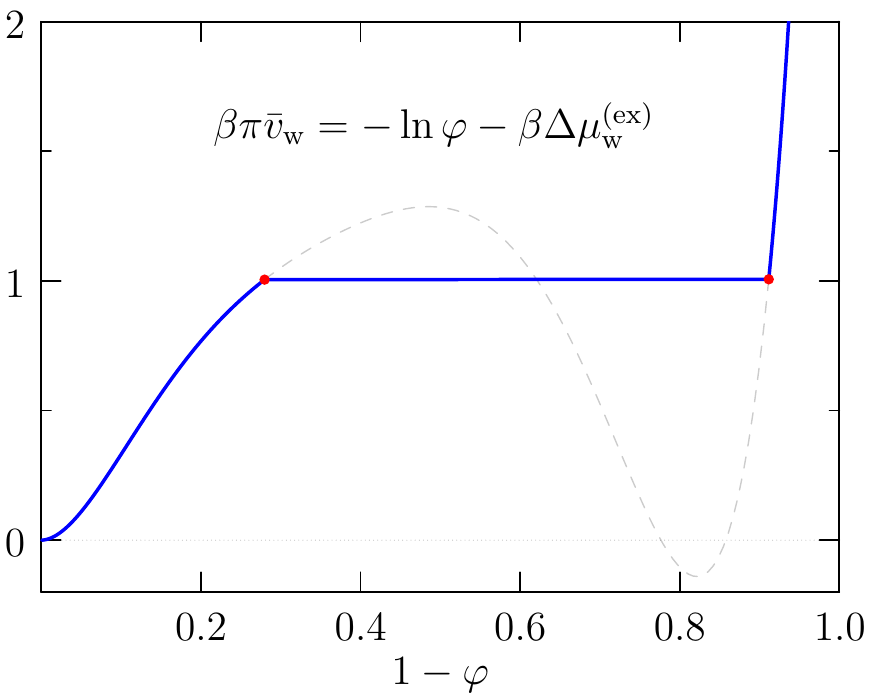}
\caption{Osmotic pressures, Eq.~\eqref{eq:pi}. The coexistence points identified
in FIG.~\ref{fig:dGmixV3} have equal osmotic pressures, as they should. The
osmotic second virial coefficient is positive.} 
\label{fig:osmoMICV5}
\end{center}
\end{figure} 

The $\chi_{\mathrm{WP}}\left(\varphi\right)$ (FIG.~\ref{fig:dGmixV3}) describes
separation of a water-poor solution from a water-rich phase. The osmotic
pressure (FIG.~\ref{fig:osmoMICV5}) further characterizes the
transition. To confirm the predicted phase separation, we
simulated coexistence of water-rich and water-poor solutions
(FIG.~\ref{fig:dGmixV3}). The two fluids did indeed coexist stably on the
simulation time scale of 20~ns, though the dynamics of the water-poor solution
are distinctly sluggish: the self-diffusion coefficient of the water in the
water-poor phase is about a 1/4th of that in the water-rich phase.

 In assessing the coexisting water-poor phase, we note that these chains are
 short and the capping groups play a significant role. Comparable
 molecular-weight PEO chains with one hydroxyl and one methoxy termination are
 pastes at low water content. Methyoxy terminated PEO chains as small as
 two-times larger than the present case form crystals with Li electrolytes at
 these temperatures.\cite{Gadjourova:2001ul} The difference between the
 predicted coexistence points and the compositions exhibited in
 FIG.~\ref{fig:dGmixV3} might be due to the assumption of ideal volumes of
 mixing. Though the excess volumes are small, they are largest in the
 interesting intermediate concentration $1-\varphi\approx 0.3$ region. This
 should receive further study. 

\section{Conclusions}
The observations here should help in formulating a defensible molecular theory of PEO
(aq) phase transitions. The interesting $\chi_{\mathrm{wp}}\left(\varphi\right)$
concentration dependence derives from long-ranged interactions. 

This analysis provides straightforward predictions of osmotic pressures, not
requiring detailed analysis of 2-body (or successive few-body) contributions.
This realization should help in studies of osmotic stress.\cite{cohen2009} 

\section{Acknowledgement} The financial support of the Gulf of Mexico Research
Initiative (Consortium for Ocean Leadership Grant SA 12-05/GoMRI-002) is
gratefully acknowledged. We thank S. J. Paddison for helpful conversations.

\section{Methods} Parallel tempering simulations, implemented\cite{chaudhari2014molecular}  within GROMACS
4.5.3,\cite{vanderSpoel:2005hz} were used to enhance the sampling at the $T$ =
300~K temperature of interest. Parallel tempering swaps were attempted at a rate
of 100/ns, which resulted in a success rates of 15-30\%. The chain molecules
were represented by the OPLS-aa force field,\cite{Jorgensen:1996vx} and the
SPC/E model was used for water.\cite{Berendsen:1987uu} Long-range electrostatic
interactions were treated in standard periodic boundary conditions using the
particle mesh Ewald method with a cutoff of 0.9~nm. The Nos\'{e}-Hoover
thermostat maintained the constant temperature and chemical bonds involving
hydrogen atoms were constrained by the LINCS algorithm. After energy minimization
and density equilibration at 300.4 K and $p$ = 1~atm, 40 replicas spanning
256-450 K, each at the same volume, were simulated for 10 ns. The pure liquid
molar volumes were $\bar{v}_{\mathrm{w}}$ = 0.0178~dm$^3$/mole,
$\bar{v}_{\mathrm{p}}$ = 0.490~dm$^3$/mole, so $M$ = 27.6. Configurations of the
$T$ = 300.4~K replica were sampled every 0.5 ps for subsequent analysis. The
packing, outer-shell and chemical terms were calculated separately using the
replica at 300.4 K. 30,000 uniformly spaced trial insertions were used to
estimate the packing term. The generalized reaction field method,\cite{Tironi:1995bo} cutoff
at 1 nm, was used to calculate the electrostatic contribution to the binding
energies. 

\bibliographystyle{achemso}

\providecommand*\mcitethebibliography{\thebibliography}
\csname @ifundefined\endcsname{endmcitethebibliography}
  {\let\endmcitethebibliography\endthebibliography}{}

\end{document}